\documentclass[twocolumn]{aastex63}

\usepackage{amsfonts,amsmath,amssymb}
\usepackage{mathrsfs}
\usepackage{nicefrac}
\usepackage{graphicx}
\usepackage{color}
\usepackage{soul} 
\usepackage{hyperref}
\usepackage{accents}
\usepackage{varioref}
\usepackage{subfigure}
\usepackage{bm}
\usepackage{algorithm}
\usepackage{algpseudocode}

\submitjournal{ApJ}

\shorttitle{DPNNet-Bayesian}
\shortauthors{Auddy et al.}
\graphicspath{{./}{figures/}}

\begin{document}

\title{Using Bayesian Deep Learning to infer Planet Mass from Gaps in Protoplanetary Disks }

\correspondingauthor{Sayantan Auddy}
\email{sauddy@iastate.edu, sayantanauddy21@gmail.com}

\author[0000-0003-3784-8913]{Sayantan Auddy}
\affiliation{Department of Physics and Astronomy, Iowa State University, Ames, IA, 50010, USA}

\author[0000-0002-0786-7307]{Ramit Dey}
\affiliation{School of Mathematical and Computational Sciences, Indian Association for the Cultivation of Science, 
Kolkata-700032, India}
\affiliation{Department of Physics and Astronomy, The University of Western Ontario, London, ON N6A 3K7, Canada }

\author[0000-0002-8597-4386]{Min-kai Lin}
\affiliation{Institute of Astronomy and Astrophysics, Academia Sinica, Taipei 10617, Taiwan}
\affiliation{
Physics Division, National Center for Theoretical Sciences, Taipei 10617, Taiwan}

\author[0000-0000-0000-0000]{Daniel Carrera}
\affiliation{Department of Physics and Astronomy, Iowa State University, Ames, IA, 50010, USA}

\author[0000-0000-0000-0000]{Jacob B. Simon}
\affiliation{Department of Physics and Astronomy, Iowa State University, Ames, IA, 50010, USA}






\begin{abstract}

Planet induced sub-structures, like annular gaps, observed in dust emission from protoplanetary disks provide a unique probe to characterize unseen young planets. While deep learning based model has an edge in characterizing the planet's properties over traditional methods, like customized simulations and empirical relations, it lacks in its ability to quantify the uncertainty associated with its predictions.
In this paper we introduce a Bayesian deep learning network ``DPNNet-Bayesian" that can predict planet mass from disk gaps and provides uncertainties associated with the prediction. A unique feature of our approach is that it can distinguish between the uncertainty associated with the deep learning architecture and uncertainty inherent in the input data due to measurement noise.
The model is trained on a data set generated from disk-planet simulations using the \textsc{fargo3d} hydrodynamics code with a newly implemented fixed grain size module and improved initial conditions. The Bayesian framework enables estimating a gauge/confidence interval over the validity of the prediction when applied to unknown observations. As a proof-of-concept we apply DPNNet-Bayesian to dust gaps observed in HL Tau. The network predicts masses of $ 86.0  \pm 5.5 M_{\Earth} $, $ 43.8 \pm 3.3 M_{\Earth} $, and $ 92.2  \pm 5.1 M_{\Earth} $ respectively, which are comparable to other studies based on specialized simulations.

\end{abstract}



\section{Introduction} \label{sec:intro}

The observed substructures, like concentric rings and gaps, from near-infrared images of protoplanetary disks (hereafter PPDs) in dust emission \citep{ALMA15,and16,Hua18a} are often interpreted as signatures of an embedded planet. However, owing to limitations in the current planet search techniques \citep{fis14,lee18}, direct detection of planets in disk is still a challenge. 

Only recently there has been measurements using gas kinematics \citep{teague2018,pinte2018,pinte19,pinte20}, that found planetary signatures, such as “kinks” due to velocity perturbations inside the observed dust gaps. Additionally, direct detection using $H \alpha$ emission of an accreting planet  \citep{kep18,2018ApJ...863L...8W,2019NatAs...3..749H} inside the cavity of the transition disk PDS-70 provides compelling evidence in favour of planetary gaps.

Thus, characterising large scale planet-induced distortions observed in PPDs give an unique opportunity to probe these unseen young planets during the epoch of their formation. It is achieved by identifying and comparing these features with theoretical models of planetary gaps generated using customized simulations \citep{Zhang2018TheInterpretation,crida06, paardekooper09, Duffell2013GAPDISK, Fung2014HOWPLANETS,Duffell2015ADISKS,Kanagawa2015MASSSTRUCTURES,ilee2020} or empirical relations \citep{kan16,lod19}.
However, owing to the increasing sample size of the observed disks and complexity of disk-planet simulations one needs to run hundreds of customized simulations for analyzing each observed system. This is too expensive and inefficient to be of practical use as a detection tool.  The alternative is to use deep learning models, trained with simulation data (generated only once), to detect/characterise exoplanets from observed data in a more efficient and accurate way.



In \cite{aud20,aud21}, hereafter Paper 1 and 2 respectively, we successfully designed deep learning models using multi-layer perceptron (MLP) or/and Convolutional Neural Network (CNN) to predict planet mass from planetary gaps in PPDs. Both models (DPNNet-1.0 and DPNNet-2.0) are end-to-end pipelines that are capable of predicting mass of an exoplanet from observed protoplanetary disk harbouring a gap inducing planet.  However, one of the main challenges with traditional deep learning models is quantifying the uncertainties associated with their predictions.
This has direct implication on how a deep learning model performs when trained on limited and/or noisy data. Furthermore,  traditional deep learning model's inability to answer ``I don't know" or ``I am not confident" is concerning, particularly, while treating \textit{out-of-training} distribution points (i.e., input data points which are outside the parameter space explored in the training dataset). Thus it is crucial to 
understand the limitations of a deep learning model by quantifying the errors associated with its predictions and identifying the source of uncertainty.

Recently some of these shortcomings are addressed by introducing probabilistic methods, such as Bayesian techniques, to deep learning architectures \citep{kendall2017uncertainties,jospin2020hands, wilson2020bayesian}. \textit{Bayesian neural network} (BNN) improves on traditional deep learning models by providing uncertainty estimates associated with the prediction of these models. BNNs are truly unique in their ability to distinguish between uncertainly caused by limitations of the model and uncertainty that is inherent to the input data. More so, these architectures can even respond to pathological out-of-training cases with a large error bar. 

For a classical neural network, one would have deterministic point estimates from a single optimized network as the output of the network \citep[e.g., see][]{aud20}.
A BNN considers aggregation of predictions (also known as marginalization) from multiple independent predictor networks  and performs a model average to get the output as a distribution. BNN is a type of stochastic neural network in which the weight are given as a distribution. It enables us to understand the uncertainty associated with the network architecture.
This is particularly powerful in constraining the error when performing a regression task to estimate parameters (such as the planet mass in our case) using deep learning models. Thus any prediction comes with a confidence interval, meeting the requirement of a scientific experiment, and one that can be compared with other methods of parameter estimation.

In this paper, we develop and introduce ``DPNNet-Bayesian", a standard MLP based model designed using Bayesian formalism to predict planet mass from disk gaps as well as quantify the uncertainty associated with the prediction. 
Our network is trained with synthetic data from explicit planet-disk simulations using \textsc{fargo3d} hydrodynamics code with a newly implemented fixed grain size module and improved initial conditions. 
 
The paper is organized as follows: In Section \ref{BBN} we give an overview of the Bayesian deep learning, its implementation and its advantage over traditional deep learning approaches. In Section \ref{sec-simulation} we describe the disk-planet hydro-simulations as well as the parameter space considered for the current study. We introduce the BBN architecture implemented in DPNNet-Bayesian in Section \ref{sec-BBN} and discuss in details the steps used to pre-process the data. In Section \ref{sec-results} we give the predictions along with the error estimates from our trained DPNNet-Bayesian, based on simulated data. In Section \ref{sec-application} as a proof-of-concept we deploy DPNNet-Bayesian to observed PPDs around HL Tau and AS 209 to infer planet masses and the associated errors. Finally, in section \ref{sec-discussion} we discuss the results and draw conclusions.

\section{Bayesian deep learning-A general overview}\label{BBN}

A Neural network architecture consists of a few thousand (or even millions) nodes/neurons that are densely interconnected. For this architecture, let

\begin{itemize}
\item  $\theta$ represent all the internal model parameters

\item $D$ represent the training data, which consists of a specified set of observed features $D_x$ (e.g., gap size) and their corresponding output labels $D_y$ (e.g., planet mass).
\end{itemize}

\noindent

Deep learning is a technique for optimizing the model parameters $\theta$ based on a training dataset $D$ \citep{goodfellow2016deep,shrestha2019review}. This is done by computing the minimum `cost' using some method of back propagation. 
Usually the cost function is defined in terms of the log likelihood of the training dataset and thus the training process can be defined as a Maximum Likelihood Estimation. This approach might suffer from over-confidence as it is deterministic in nature. 
An alternative to this is to implement the use of stochastic components in the neural networks. This can account for the uncertainties in the determination of the network parameters $\theta$ that finally shows up as an estimated error in the predictions made by a deep learning architecture \citep{jospin2020hands}. For these type of neural nets, the stochastic elements can be introduced in terms of stochastic model weights or activation. Thus, instead of a single deterministic output from the traditional deep learning model, for the stochastic networks a marginalization of the output is considered based on some statistical principle \citep{wilson2020bayesian}.

In order to design a predictive model based on BNN, one starts with a suitable choice of a deep learning architecture. The next step is to choose a systematic prescription in order to introduce the stochastic elements in the network architecture. This can be modeled in terms of some possible prior distribution of the model parameters, $p(\theta)$.

From Bayes' theorem we get
\begin{align}
p(\theta | D)\propto p(D_y|D_x, \theta) p(\theta)
\end{align}
where $D_y$ denotes the training labels and $D_x$ are the training features as mentioned earlier. Bayesian posterior distribution, $p(\theta|D)$, depicts the model parameters as a probabilistic distribution rather than some deterministic value. 

The Bayesian approach allows us to compute model averages by essentially integrating over models, parameterized by the model parameters $\theta$. The distribution of possible values of $y$ (e.g., planet mass) given some input $x$ (e.g., dust gap width) is given by 

\begin{align} \label{output}
p(y|x,D)=\int_{\theta}p(y|x,\theta')p(\theta'|D)d\theta',
\end{align}
In equation \eqref{output}, the variance of $p(y|x,\theta')$ captures the uncertainty due to measurement noise of the input data while the variance of $p(\theta'|D)$ corresponds to the uncertainty associated with the neural network architecture.

The classical training would correspond to, $p(\theta | D)$  replaced by a constant value of $\theta$, but in the Bayesian case this would typically be a probability distribution with some finite variance \citep{wilson2020bayesian, abd21}. Also, computing $p(\theta|D)$ from the training data is one of the most challenging and difficult aspects of the BNN architectures due to the large size of the parameter space. This is further discussed in the following sections.

\subsection{The variational inference approach to Bayesian deep learning}
\label{VI}
Due to the large dimensionality of the sampling space, it can get quite challenging to merge the concepts of a neural network with the Bayesian approach. As the posterior distribution is extremely complex, directly estimating $p(\theta|D)$ to perform the integral in Equation \eqref{output} is not always possible.
Different methods, involving various approximations, have been implemented to do this. One of the most popular and direct method for sampling the posterior distribution, $p(\theta|D)$, is using Markov Chain Monte Carlo (MCMC) \citep{bardenet}. However, MCMC often has issues with scalability owing to large model sizes and it can be computationally expensive as well.

Instead of sampling from the posterior directly, the method of variational inference can be used as an alternative \citep{blei2017variational, Mohan:2022oac}. 
A distribution (referred to as the variational distribution), parametrized by an arbitrary set of parameters $\phi$, is used to approximate the posterior. The parameter $\phi$ is learned in a way that the distribution $q_{\phi}(\theta)$ becomes close to the exact posterior $p(\theta|D)$. The closeness between the posterior and $q_{\phi}(\theta)$ is quantified in terms of the KL-divergence \citep{zhang2018advances,jospin2020hands}. More precisely, the formalism of \textit{evidence lower bound} (ELBO) is used for this purpose. 
\begin{align}
ELBO=\int_\theta q_{\phi}(\theta')\log \left( {p(\theta',D)\over q_{\phi}(\theta')}\right)d\theta'
\end{align}
Minimizing the ELBO is equivalent to minimizing the KL-divergence as $ELBO=\log (p(D))-D_{KL}(q_{\phi}||p)$, where $\log (p(D))$ depends on the prior and $D_{KL}$ is the KL-divergence.

\begin{figure*}[ht!]
\centering
\includegraphics[scale=.35]{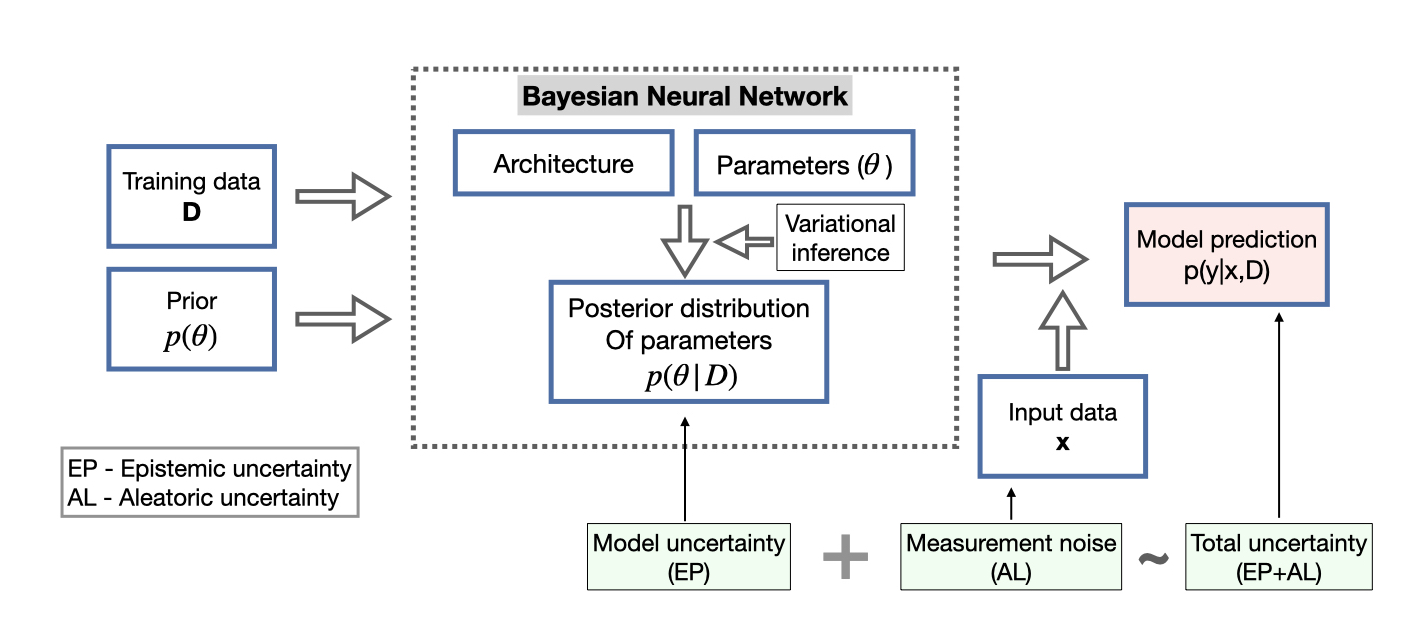}
\caption{This schematic diagram shows the workflow of a typical Bayesian Neural Network along with how the uncertainty/errors propagate.
}\label{fig:dpnnet_bayesian}
\end{figure*}

\subsection{Features and advantages of BNN}
\label{add_bnn}

Once the ELBO function is optimized and thus the variational distribution is learned/trained we can use it to predict the labels of new (unseen) observations. The predictive posterior distribution  $p(y^*|x^*,D)$  is given by integrating out the model/variational parameters (trained using the dataset $D$):

\begin{align} \label{output_2}
p(y^*|x^*, D)=\int_{\theta}p(y^*|x^*,\theta')p(\theta'|D)d\theta',
\end{align}
where $(x^*, y^*)$ represents the features and labels respectively and the super-script ``$*$" denotes unseen data.
In order to obtain this predictive distribution we use Monte Carlo sampling. First, we sample from the distribution of model parameters $N$ times,

\begin{equation}\label{eqn:theta_j}
    \theta^j \sim p(\theta|D)
    \;\;\;\;\;\;\;\;\;\;j\in\{1, 2, ..., N\}.
\end{equation}

\noindent
Then $x_{i}$ is sampled $n$ times from the distribution of the input features, whose width is determined by the errors associated with the feature itself, to estimate the predictive distribution
$p(y_i^*|x_i^*,\theta^j)$.
For $N \times n$ iterations we get
\begin{align}\label{uncertianty}
    p(y^*|x^*,D) \approx \sum_{i=1}^n \sum_{j=1}^N p(y_{i}^{*}|x_{i}^{*},\theta^{j})p(\theta^{j}|D).
\end{align}

\noindent

 The error/uncertainty in the predictions of BNN can be broadly classified into two categories: epistemic and aleatoric uncertainty  \citep[][]{abd21,hullermeier2021aleatoric}. 
``Aleatoric'' uncertainty (AL) represents the uncertainty inherent in the data which is given as the input for the model (e.g measurement noise). In contrast ``Epistemic'' uncertainty (EP) represents limitations of the ML based model which usually stems from lack of information about the system. Taking a larger dataset for training the model (spanning a larger parameter space) or improving the ML architecture can typically reduce the Epistemic uncertainty. On the other hand as the aleatoric uncertainty is intrinsic to measurement noise of the input data it cannot be reduced by solely improving the ML model. Both these uncertainties propagate to the output to give the net predictive uncertainty \citep[][]{abd21} as shown in Figure (\ref{fig:dpnnet_bayesian}).

We can measure the epistemic uncertainty, $\sigma_{\rm EP}$, from the variance of the posterior distribution of the model parameters, $p(\theta^j | D)$, given as
\begin{align}
\sigma_{\rm EP}^2 \sim \text{Variance of } p(\theta'|D).
\end{align}
$\sigma_{\rm EP}$ quantifies the intrinsic uncertainty associated with the model. It enables us to understand the correlation between the model performance, the quantity/quality of the data and the parameters of the ML architecture. Thus, monitoring $\sigma_{\rm EP}$ can give insights on improving the overall ML model.

The aleatoric uncertainty, $\sigma_{\rm AL}$, is obtained from the variance of $p(y^*_i|x^*_i,\theta^j)$, given as
\begin{align}
\sigma_{\rm AL}^2  \sim  \text{Variance of } p(y|x,\theta).
\end{align}
$\sigma_{\rm AL}$ quantifies the statistical uncertainty/error associated with various measurements that goes as input features to the neural net at the training stage as well as when the model is deployed for real world application. Knowing $\sigma_{\rm AL}$, one can monitor the errors in the prediction of the model as the measurement noise of the input features changes.
The advantage of knowing both $\sigma_{\rm EP}$ and $\sigma_{\rm AL}$ is that, while testing the model, one can trace the source of uncertainty and improve the model/dataset accordingly.

\section{Disk-Planet Simulations}\label{sec-simulation}
We model the structural variations of a dusty protoplanetary disk due to an embedded gap-opening planet using \textsc{fargo3d} hydrodynamic simulation code \citep{Benitez-Llambay2016FARGO3D:CODEb}. We update the public version of the code by parametrizing the dust gas interaction with constant particle size, instead of fixed Stokes number. This includes a generalization of the steady state profiles derived by \citet{llambay19}, which are presented in Appendix \ref{append1}. We regenerate the dataset with the updated simulations to train the BBN model. The simulation setup and the parameter space is discussed in the next section.

\subsection{Protoplanetary Disk Setup}
We model a simple 2D razor-thin disk with an embedded planet of mass $M_{\rm P}$ located at $R=R_{\rm 0}$. The planet is on a Keplerian orbit around a central star of mass $M_{*}$ and is not migrating. The unit of time is the planet's orbital period $P_{0} = 2\pi / \Omega_{\rm K0}$, where $\Omega_{\rm K0}$ is the Keplerian frequency. The sub-script $0$ denotes evaluation at the planet's location $R_{0}$.

The disk's initial gas surface density profile is given as 

\begin{figure*}[ht!]
\centering
\includegraphics[scale=.70,]{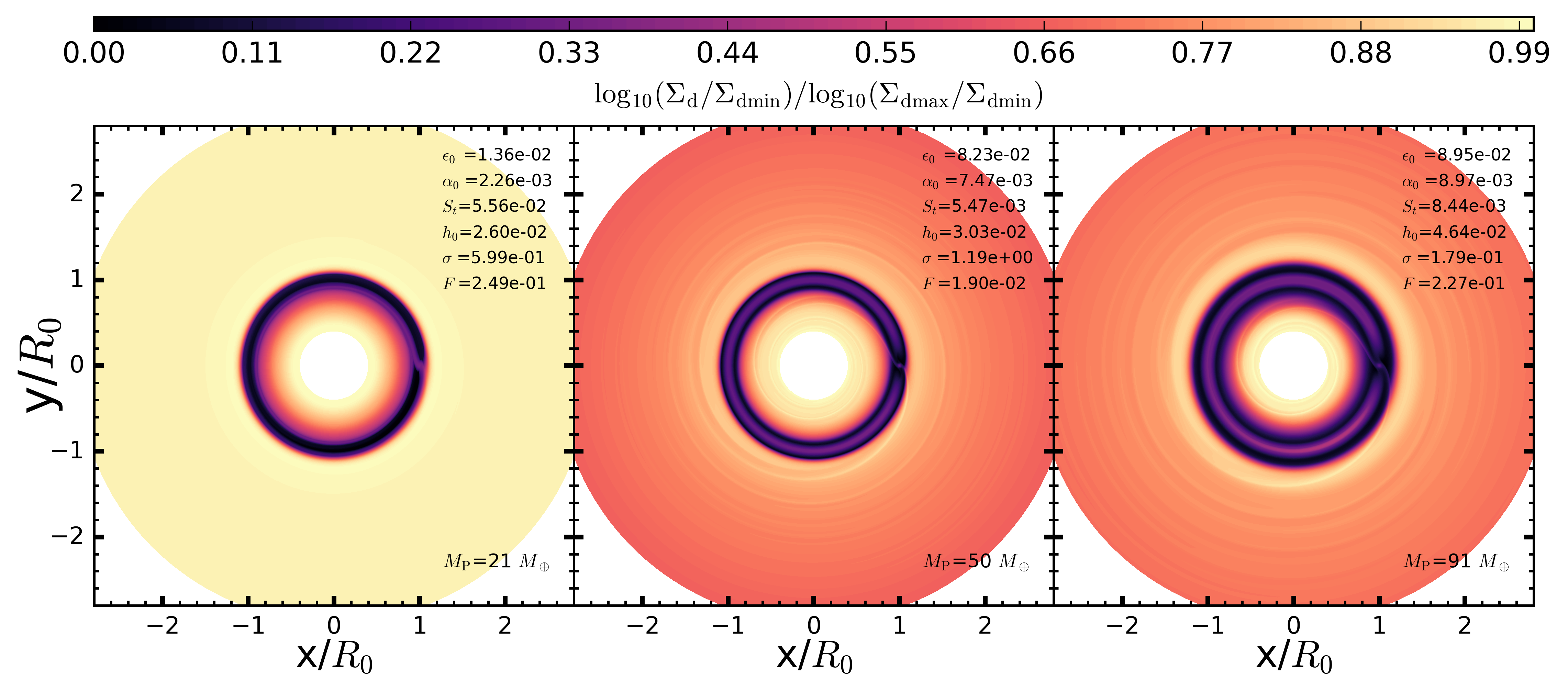}
\includegraphics[scale=.63,]{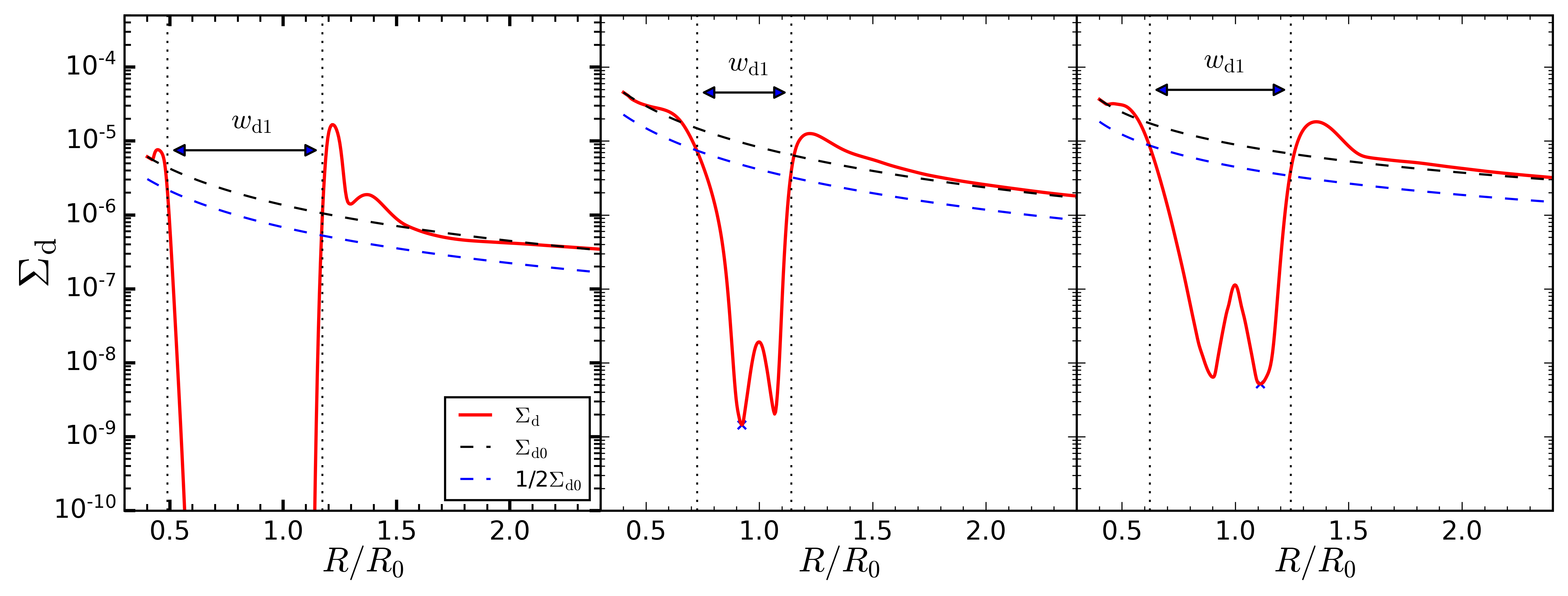}

\caption{The upper panel is the normalized dust surface density distribution of disks with different initial conditions and  for different planet masses at $ 3000 P_{\rm 0}$. The disk parameters are indicated on the top right of each plot along with the planet mass in the bottom right. The lower panel is the radial profile of the azimuthally averaged surface density at the beginning of the simulation ($\Sigma_{\rm d0}$; dashed black) and at the end of the simulation after the gap is formed ($\Sigma_{\rm d}$; solid red). The horizontal arrow marks the dust gap width. The gap width is the distance between the inner and the outer edge of the gap (vertical dotted lines) where $\Sigma_{\rm d}(R)$ reaches $50$ \%  of the initial surface density $\Sigma_{\rm d0}$. The cross indicates the minimum surface density $\Sigma_{\rm dmin}$. }\label{fig:Sample_data}
\end{figure*}

\begin{equation}
\Sigma_{\rm g}(R) = \Sigma_{\rm g0} \left(\frac{r}{r_{0}}\right)^{-\sigma},   
\end{equation}
where $\sigma$ is the exponent of the surface density profile and $\Sigma_{\rm g0} = 10^{-4}$ is a constant in code unit, which is arbitrary for non-self-gravitating simulations without orbital migration. For reference, however, this surface density scale corresponds to Toomre parameters $\mathcal{Q} \geq 80$ for all of our disk models, which is sufficient for the disk to be gravitationally stable \citep{toomre64}.
The disk is flared with a constant flaring index F, such that the aspect ratio varies as

\begin{equation}
    h = \frac{H}{R} = h_{0} \left( \frac{R}{R_{0}} \right)^{F},
\end{equation}
where $H$ is the disk scale height and $h_{0}$ is the aspect ratio at $R=R_{0}$. The disk is locally isothermal with a sound-speed profile of $c_{s}(R) = h_{0}R\Omega_{\rm K}$. To mimic the disks turbulence we include viscous forces in the gas disk using the standard $\alpha$-prescription \citep{shakura73}, where the kinematic viscosity $\nu = \alpha c_{\rm s}^2/\Omega_{\rm K}$ and $\alpha$ is set by Equation \ref{alpha}.

We consider a single population of dust particles modeled as a pressureless fluid \citep{jacquet11}. The dust particle size is parameterized using Stokes numbers $S_{\rm} \equiv t_{\rm s}\Omega_{\rm K}$, where  $t_{\rm s}$ is the stopping time  characterizing the strength of the dust-gas coupling \citep[][]{weiden77}.  The initial dust surface density is  $\Sigma_{\rm d} = \epsilon \Sigma_{\rm g}$, where $\epsilon$ is the dust-to-gas ratio set by Equation (\ref{epsilon_new}). The dust back reaction onto the gas is included in the simulations.

\subsection{Simulation Setup and Parameter Space}\label{paramspace}
We run \textsc{fargo3d} planet-disk simulations on Graphical Procession Units (GPUs). Our computational domain is $R \in [0.4, 3.0]R_0$ and $\phi\in[0, 2\pi]$, and are discretized with 512 $\times$ 512 logarithmically-spaced cells. The radial and the azimuthal boundaries are set to their initial equilibrium solutions given in equations \ref{vgr} - \ref{vdphi}. The planet mass is allowed to grow slowly over first $100 P_{\rm 0}$. We implement the wave-killing \citep{valborro06} module to damp the planet generated waves to zero near boundaries to minimize reflection and interference with disk morphology.  

We target super-Earth to Saturn-sized planets with masses $M_{\rm P} \in [8,120] M_{\earth}$ around a 1 $M_{\odot}$ central star. We model the disk for a range of values for disk aspect ratio $h_{\rm 0} \in [0.025,0.1]$, viscosity parameters $\alpha_{\rm 0} \in [10^{-4}, 10^{-2}]$, flaring index $F \in [0,0.25]$, and power-law slope of the surface density profile $\sigma \in [0.05,1.2]$. In each disk we consider a single dust species of fixed size characterized by Stokes number $S_{\rm t} \in [10^{-3}, 10^{-1}]$ and abundance $\epsilon_{\rm 0} \in [0.01,0.1]$. This range of $S_{\rm t}$ corresponds to particle radii ranging from $\sim 1$ mm to $\sim 10$ cm, for typical gas surface density of $100 \rm g cm^{-2}$ and internal grain density of $1 \rm g cm^{-3}$.

\section{DPNNet-Bayesian}\label{sec-BBN}
In this section we introduce the Bayesian implementation of the DPNNet model. 

\subsection{Network architecture}\label{architecture}
The DPNNet model as presented in \cite{aud20} was based on artificial neural networks with a linear layer in the end for performing regression. In the Bayesian adaption of DPNNet the front end of the model includes a MLP layer with 256 units of neuron having a ReLu activation function (this part of the network is the same as DPNNet-1.0). This is followed by a fully connected variational dense layer having 128 neurons, in which the BBN algorithms are implemented to probe the model uncertainty. We use the method of variational inference, as described in Section \ref{VI}, to account for the stochasticity of the model parameters. Each model parameter is treated as a distribution having some variance. The stochastic sampling from this variational distribution gives an estimate of the model uncertainty. Furthermore, we use a trainable prior distribution for the parameter space, $p(\theta)$. This is used to determine the approximate posterior distribution of the model weights using variational inference.

\subsection{Data acquisition and pre-processing}
For training and testing the model, we produce the new dataset following the same procedure as used in Paper 1. We generate an uniform random distribution of the parameters, each centered within the sampling interval, using the Latin hypercube sampling method \citep{mck79,ima81}. We run a total of 1800 simulations with a range of input parameters described in section (\ref{paramspace}). From each simulation we get the dust and the gas surface density maps of the PPD with the embedded planet. Figure \ref{fig:Sample_data}, shows the sample images of dust surface density distribution and the corresponding radial profile of the azimuthally averaged surface density from some of the simulations. These improved disk-planet simulations (with fixed grain sizes and initial equilibrium disk setup) result in much physical model for dust-gas interaction compared to Paper 1 and Paper 2.

Following Paper 1 and \cite{kan16}, we define the dust gap width as 
\begin{equation}
    w = \frac{R_{\rm d, out} - R_{\rm d, in}}{R_{\rm 0}}
\end{equation}
where $R_{\rm d, in}$ and $R_{\rm d, out}$ are the inner and the outer edge of the gap where the dust surface density $\Sigma_{d}$ equals a predefined threshold fraction of $\Sigma_{\rm d0}$. For consistency with Paper 1, we adopt a threshold fraction of 1/2 of the initial surface. We acknowledge this definition of $w_{\rm d}$ has its limitations as it fails for shallow gaps (for low mass planets), when gap depth is less than the threshold function. However, this does not limit the model application as DPNNet-Bayesian can be retrained for any threshold function or other definition of the gap width.

In the current paper we only consider simulations with either one dust gap or with two dust gaps. We remove models with $>2$ dust gaps as the number of samples are not adequate to effectively train the model. To build our feature space we simply define two dust gap widths for each simulations, $w_{\rm d1}$ and $w_{\rm d2}$, but set $w_{\rm d2} = 0$ if only single dust gap is formed. 

The initial dataset consists of 1800 simulations. We pre-process the data by eliminating runs that do not open up detectable axis symmetric gaps or have more than two gaps. After the initial filtering we have data from 1098 simulations. Each simulation is characterised by its input parameters ($M_{\rm P}$, $\alpha_{\rm 0}$, $h_{0}$,$\sigma$,$F$,$S_{\rm t}$ $\epsilon_{\rm 0}$) and the associated gap width(s) in dust ($w_{\rm d1}$,$w_{\rm d2}$) and gas ($w_{\rm g}$). In this paper we will not use gas gap width as an input parameter for training the model since these are not directly measured from observations.

Once the gap widths are measured, we assign each simulation a set of feature variables ($w_{\rm d1}$,$w_{\rm d2}$, $\epsilon_{\rm 0},\alpha_{\rm 0}, S_{\rm t}, h_{\rm 0}, \sigma, F$) that would go as the input for our model and label the planet mass, $M_{\rm P}$, as the target variable. The data for each feature variable is then normalized using the standard (z-score) scaling by subtracting the mean and scaling it by the standard deviation.

\subsection{Training Bayesian-DPNNet}
The Bayesian DPNNet is implemented using Google TensorFlow Probability \citep{tensorflow}, which is an open source platform that integrates probabilistic methods with deep neural networks. 
We split the data randomly into two blocks, a training set consisting of 80 \% and a testing set consisting of 20 \% of the simulation data. Furthermore, for model validation we use 20\% of the training dataset. We use the Adam optimizer with a learning rate of $0.001$.
The model is trained for $\sim 2200$ epochs and the early stopping callback algorithm is used to prevent the model from over-fitting.

\section{Result }\label{sec-results}

Once the model is trained it is ready to be deployed to predict planet mass from observed disk features. We test the networks performance on the test dataset which the model has not been exposed to. This enable us to quantify the networks accuracy in predicting planet mass from unseen data. The test dataset comprises of all the feature variables along with the target variable (the true value of the planet mass). The feature variables ($w_{\rm d1}$,$w_{\rm d2}$, $\epsilon_{\rm 0},\alpha_{\rm 0}, S_{\rm t}, h_{\rm 0}, \sigma, F$)  are normalized with the same scaling that the network was trained on. The trained DPNNet-Bayesian network takes the feature variables as input and predicts the corresponding planet mass. Unlike the standard MLP model, for the same set of input feature variables, DPNNet-Bayesian gives the predicted output as a distribution. 

\begin{figure}[ht]
\centering
\includegraphics[scale=.65]{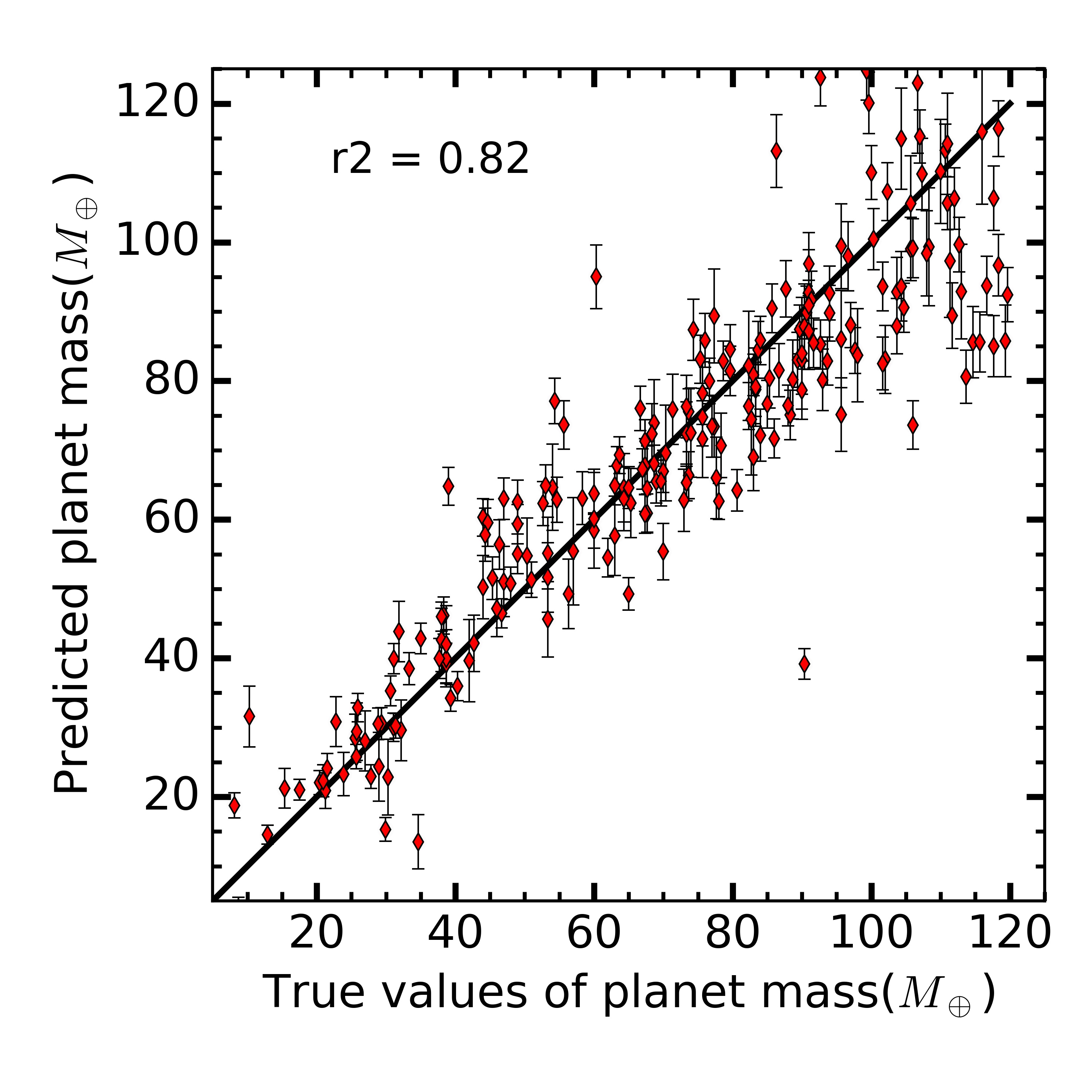}
\caption{Correlation between the simulated and the mass predicted mass in the units of $M_{\earth}$. The vertical error bar gives the standard deviation associated with each predicted value obtained using the BNN. The $r2$ score indicate the
goodness of fit. 
}\label{fig:correlation_plot}
\end{figure}

\begin{figure*}[ht!]
\centering
\includegraphics[scale=.60]{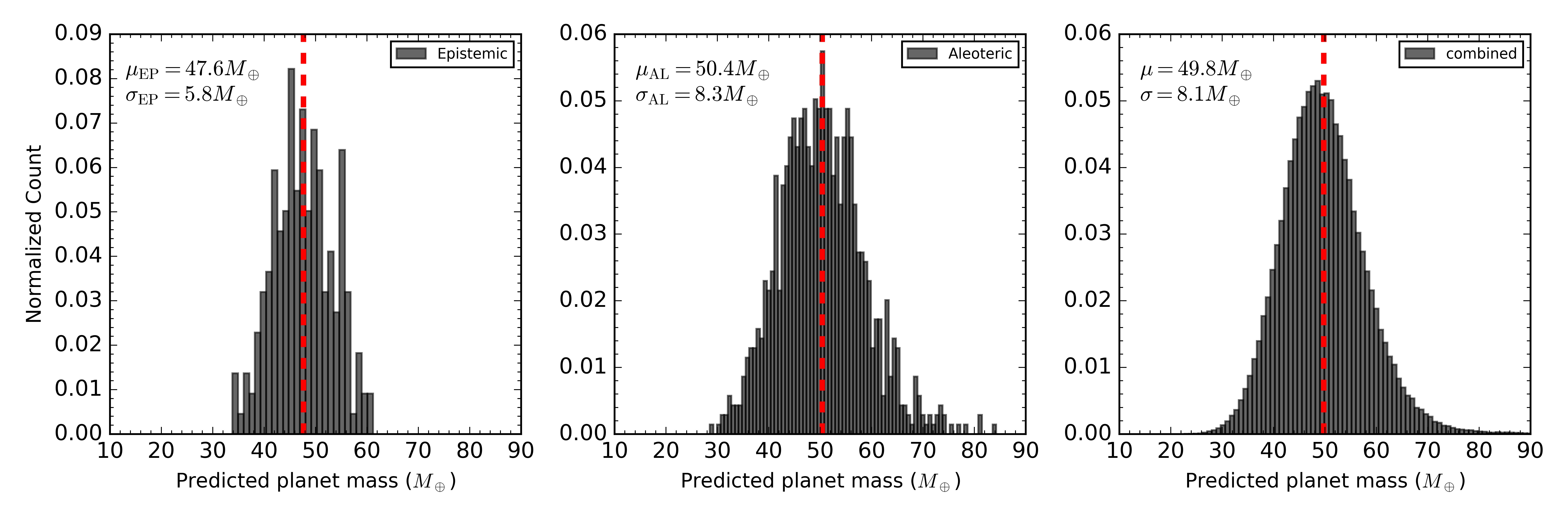}
\caption{Distribution of the predicted planet masses for sample disk features selected randomly from the test dataset. Left Panel: captures the epistemic uncertainties associated with DPNNet due to lack of data and/or variations of the model parameters. Middle Panel: represents the aleatoric uncertainties which arises due to errors/variations inherent in the data. Right Panel: captures the combined uncertainty in  the  predicted  planet  mass  due  to  both  the epistemic 
and aleatoric  uncertainties. The red dashed line represents the mean planet mass in each of the plots. }\label{fig:uncertainty}
\end{figure*}

Figure \ref{fig:correlation_plot} shows the correlation between the simulated and the predicted planet mass in Earth-mass ($M_{\earth}$) units for all the test samples. For each set of input variables we conduct the experiment (i.e., deploy the trained network) 200 times to obtain the marginalization of the predicted output. 
Thus we obtain a distribution for the predicted planet mass (as demonstrated in figure (\ref{fig:uncertainty})) capturing the uncertainties associated with the parameters of the trained model. The error bar gives the standard deviation associated with each predicted value. The predicted value lies along the black line indicating a strong correlation with a $r2$-score of $\rm r_b= 0.82$.

\subsection{Error Analysis using BNN}
As mentioned before, the Bayesian approach offers a way to quantify and understand the uncertainties associated with predictions from the deep neural networks. 
As an example, in Figure \ref{fig:uncertainty} we demonstrate the different uncertainty estimates associated with prediction from DPNNet-Bayesian for sample disk features selected randomly from the test dataset. We pick a disk-planet simulation with parameters/features $\epsilon_{\rm 0}=0.04,\alpha_{\rm 0}=0.008 , S_{\rm t}=0.08, h_{\rm 0}=0.03, \sigma =1.11, $ and $F=0.21$ and measured dust gaps widths $w_{\rm d1}$ =0.74 , $w_{\rm d2}$= 0. Since these are simulated data the true value of the planet mass, $M_{\rm P} = 46 M_{\earth}$, is known. To capture the epistemic uncertainty we deploy the trained model $N$ times for these fixed set of disk features. We set $N = 200$, with the $i$ index fixed, in Equation (\ref{uncertianty}) and for each iteration the model predicts a unique and independent value. This results in a distribution of predicted planet mass as seen in the left panel in Figure \ref{fig:uncertainty}. The mean ($\mu_{\rm EP} =  47.6 M_{\earth}$) shown with the red dotted line, is taken as the final predicted value while the standard deviation ($\sigma_{\rm EP} = 5.8 M_{\earth} $) represents the uncertainty associated with the trained model.

The measured disk features are also prone to statistical errors due to measurement techniques and limitations in the observed data. While parameters in simulated models are well constrained, that is not the case for observed disk data. In order to simulate the uncertainties associated with the measured disk parameters we use Monte Carlo analysis. Instead of using fixed true values for the disk parameters we sample them randomly $n$ times from their respective standard normal distribution with mean set to the true value and standard deviation equal to $1 \%$ spread for each of the parameters. We set $n=1000$, with $j$ fixed, in Equation (\ref{uncertianty}) such that for each set of sampled input parameters the model predicts a planet mass. The center panel of the Figure \ref{fig:uncertainty} demonstrates the aleotoric uncertainty $ \sigma_{\rm AL}= 8.3 M_{\earth} $ corresponding to the predicted planet mass due to the measurement noise of the input disk parameters. 

The right most panel in Figure \ref{fig:uncertainty} captures the total uncertainty in the predicted planet mass due to both the epistemic and aleatoric uncertainties . We sample the disk parameters randomly, $n=1000$ times, from their standard normal distribution and for each draw we deploy the model  $N = 200$ times. This results in distribution of predicted planet mass with mean $\mu=49.8 M_{\earth}$ and standard deviation $\sigma=8.1 M_{\earth}$.

\begin{figure*}[ht!]
\centering
\includegraphics[scale=.65,]{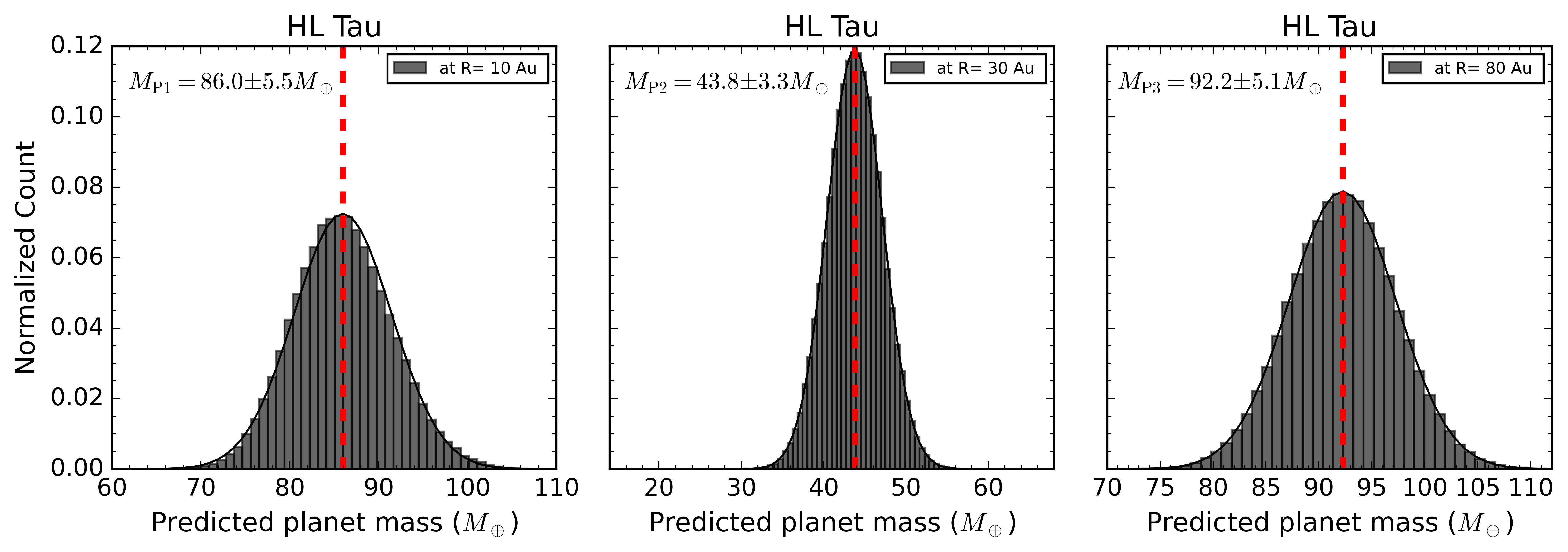}
\includegraphics[scale=.65,]{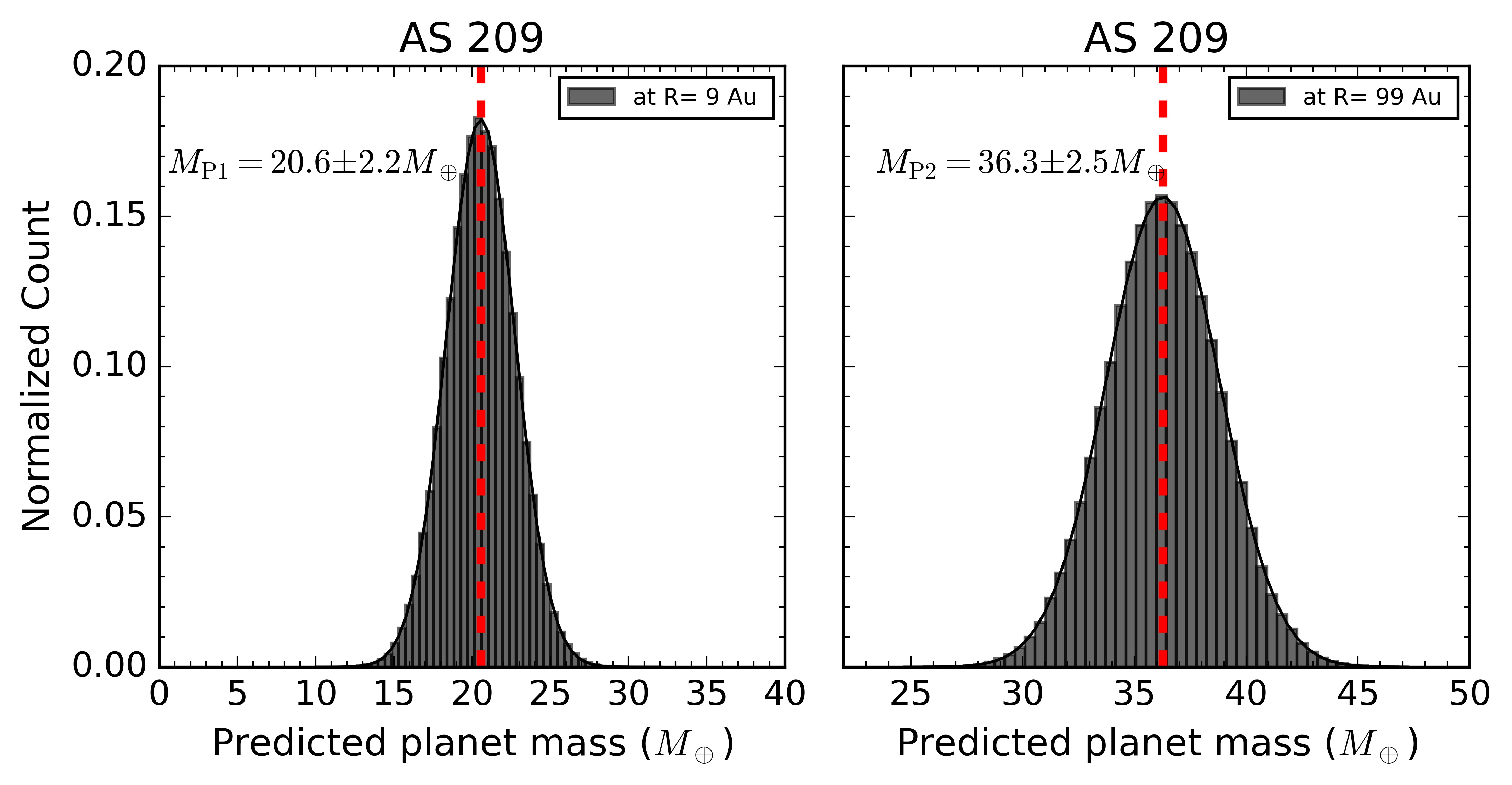}
\caption{Top Panel: Distribution of the predicted planet mass at three identified gaps in HL Tau obtained using DPNNet-Bayesian. The mean values along with the standard deviations of the predicted planet masses at 10 au, 30 au, and 80 au are  $ 86.0  \pm 5.5 M_{\Earth} $, $ 43.8 \pm 3.3 M_{\Earth} $, and $ 92.2  \pm 5.1 M_{\Earth} $ respectively. Bottom Panel: shows the distribution of predicted masses for the two identified gaps in AS 209 for the given disk parameters and gap widths. The mean predicted planet masses along with the standard deviation are $20.6 \pm 2.2 M_{\Earth} $ and  $ 36.3  \pm 2.5 M_{\Earth}$ in $9$ au and $99$ au respectively.}\label{fig:HL-Tau_mass}
\end{figure*}

\section{Application to Observation} \label{sec-application}
As a first test of our newly improved DPNNet-Bayesian model, we deploy it to estimate planet masses from the observed dust gaps in the protoplanetary disks around HL Tau and AS 209. The results are elaborated and discussed below. \\

\subsection{HL Tau}

HL Tau is a well studied system, which shows distinct axisymmetric - possibly planet induced - gaps in the dust emission \citep{ALMA15}. Several studies involving hydrodynamic simulations \citep{Dong2015OBSERVATIONALDISKS,Jin2016MODELINGINTERACTIONS,Dipierro2015OnTau} suggest that each of the gaps are induced by a single planet. With the assumption that the observed gaps are all planet induced, we use the properties of the observed gaps as input features to our BNN model to predict the planet mass. DPNNet-Bayesian predicts a distribution for the planet masses in each of the gaps. For the three identified gaps at R = 10 au, 30 au and 80 au, we obtain the gap widths ($w_{\rm d1} = 0.81, 0.23, \, \rm {and} \, 0.29$) and adopt the disk aspect-ratios of $h_{\rm 0} = 0.05,0.07, \, \rm {and} \, 0.10$ respectively from \cite{Kanagawa2015FormationRotation,kan16}. HL Tau is mostly laminar with viscosity $\alpha_{\rm 0} = 10^{-3}$ \citep{Dipierro2015OnTau} and a central star mass of $M_{*}=1 M_{\odot}$. We set a canonical value of the dust-to-gas ratio $\epsilon_{\rm 0} = 10^{-2}$ \citep{ALMA15,Dong2015OBSERVATIONALDISKS} and adopt a mean Stokes number $S_{\rm t} = 0.005$ \citep{Dipierro2015OnTau} on all three gaps. Since the uncertainty associated with the input parameters is not well constrained across the literature, we take a conservative approximation and consider $1-5 \%$ variation in the gap widths only.
The top panel of Figure \ref{fig:HL-Tau_mass} shows the distribution of predicted masses in all the three identified gaps for the given disk parameters and gap widths. The mean value along with the standard deviation of the predicted planet masses at 10 au, 30 au, and 80 au are $ 86.0  \pm 5.5 M_{\Earth} $, $ 43.8 \pm 3.3 M_{\Earth} $, and $ 92.2  \pm 5.1 M_{\Earth} $ respectively. This standard deviation captures the total uncertainty, which includes the uncertainty associated with the BNN architecture ($\sigma_{\rm EP}$ estimated as $ 4.0 M_{\Earth} $, $ 3.3 M_{\Earth} $, and $ 5.0 M_{\Earth}$ respectively for the three gaps) along with the aleatoric uncertainty. For the assumed variation of $1-5 \%$ in the gap widths, $\sigma_{\rm AL}$ is comparable to $\sigma_{\rm EP}$. However, $\sigma_{\rm AL}$ can vary depending on the noise in the input data.  These estimates are mostly consistent with the inferred planet mass from direct numerical simulations like \cite{Dong2015OBSERVATIONALDISKS,Jin2016MODELINGINTERACTIONS,Dipierro2015OnTau} and DPNNet-1.0 in Paper 1.

\subsection{AS 209}
AS 209 is another well studied system hosting multiple gaps at 9, 24, 35, 61, 90, 105 and 137 au \citep{Hua18b}. However, for simplicity we only consider the two most prominent gaps at R = 9 and 99 au. We adopt the disk parameters and dust gap widths from \cite{Zhang2018TheInterpretation} and use a range of values for dust abundance, particle size and disk viscosity which are within our training parameter space. For examples, we set $\alpha_{\rm 0} = 10 ^{-4}$, $S_{\rm t} = 156.9 \times 10^{-4}$ and surface density profile $\sigma = 1$  \citep{fed18} for both the gaps. Next, for the innermost gap at $R= 9$ au, we assign values to the other parameters as $w_{\rm d1} = 0.42$, $h_{\rm 0}= 0.04$ and $\epsilon_{\rm 0} =0.012$. Similarly for the outer gaps at $R=99$ au we select $w_{\rm d1} = 0.31$, $h_{\rm 0}= 0.08$ and $\epsilon_{\rm 0} =0.017$. Once again we consider $1-5 \%$ variation in the gap widths as a conservative approximation to capture the uncertainty associated with its measurement. The bottom panel of Figure \ref{fig:HL-Tau_mass} shows the distribution of predicted masses for the two identified gaps in AS 209 for the given disk parameters and gap widths. The mean predicted planet masses along with the standard deviation are $20.6 \pm 2.2 M_{\Earth} $ and  $ 36.3  \pm 2.5 M_{\Earth}$ at $9$ au and $99$ au respectively. The standard deviation captures the total uncertainty which includes the epistemic uncertainty ($\sigma_{\rm EP}$ is  $ 2.0 M_{\Earth} $ and $ 2.1. M_{\Earth}$ for the two gaps respectively) along with the aleatoric uncertainty.  $\sigma_{\rm AL}$ is comparable to $\sigma_{\rm EP}$ for the assumed variation of $1-5 \%$ in the gap widths. However, $\sigma_{\rm AL}$ can vary depending on the noise in the input data. These estimates particularly at $R=9$au are lower compared to \cite{Zhang2018TheInterpretation}1. The discrepancy is partially due to differences in the way the gap widths are measured, the uncertainty in the adopted disk parameters and simulation setup.

\section{Discussion and Conclusion}\label{sec-discussion}

In this paper we introduced DPNNet-Bayesian, a BNN based model, which characterizes mass of unseen exoplanets from observed PPDs. Compared to its predecessor DPNNet-1.0, the Bayesian approach has the following advantages: Firstly, it quantifies the errors associated with the prediction of the network and trace the source of the uncertainty. Secondly, it gives a framework to understand various regularization techniques used in traditional deep learning approaches \citep{jospin2020hands}. Finally, compared to traditional deep learning models, where the weights are initialized by some implicit prior, the Bayesian framework accounts for this prior distribution explicitly. 

For a BNN, one trains a distribution/ensemble of deep learning models and then based on the Bayesian formalism, performs model averaging/marginalization of the output. It is to be noted that the BNN approach is fundamentally different from ``deep ensembles", where different initialization are used to train a set of network and then the outputs are combined/optimized.
More so in the BNN approach, a weighted averaging is done by taking into account the posterior of the network parameters as weight. This is different from the deep ensembles where the averaging is directly done with the output of the network.

Furthermore, the error/uncertainty associated with the output of a BNN model is different from the standard metrics, like root mean square error (RMSE) and/or mean absolute error (MAE), used in traditional deep learning models. Both MAE and RMSE are measure of the error ($|M_{\rm P, simulation}- M_{\rm P, predicted}|$) when the model is deployed on a validation/testing dataset for which the ``true" target values are known. However they do not necessarily quantify the error accurately (or provide a confidence interval) when applied to a unknown dataset for which the target values are not known. This can be problematic as there is no way of knowing how accurate the prediction is and one needs to trust on the validation/testing error as an indication of the accuracy of the prediction. Since a BBN model works on the principle of marginalization rather than optimisation, the output is obtained by sampling from a posterior distribution, thus, providing the necessary confidence interval.

The performance of any ML based model depends on the quality and/or quantity of the training dataset. The Bayesian approach helps us to understand the correlation between the quality/quantity of the dataset and the model performance by monitoring the change in model uncertainty.  The knowledge of the epistemic and aleatoric uncertainties enables the model to learn from a smaller dataset by regulating over-fitting or under-fitting. Thus the use of BNN architecture plays a pivotal role in improving the training process as well as determine the optimal deep learning architecture.\\

We conclude by highlighting the key results:

\begin{itemize}
\item DPNNet-Bayesian estimates the planet mass from the gap width observed in dust emission. It takes as input dust properties (abundance, Stokes numbers), gas disk properties (aspect-ratio, surface density profile, viscosity), along with dust gap widths from observations and predicts the embedded planet mass.

\item For training DPNNet-Bayesian we generated a dataset from disk-planet simulations using the \textsc{fargo3d} hydrodynamics code with a newly implemented fixed grain size module and improved initial conditions.
The trained model is tested on the simulated dataset in order to illustrate its applicability. The predicted mass is closely correlated with the `true' value of the simulated planet mass. 

\item Compared to DPNNet 1.0 \citep{aud20},  the DPNNet-Bayesian architecture additionally estimates the error associated with the prediction of the model.
It further distinguishes between the uncertainty pertaining to the deep learning architecture (Epistemic uncertainty) and the errors associated with the input variables due to measurement noise (Aleatoric uncertainty).

\item  We deploy DPNNet-Bayesian to dust gaps observed in the protoplanetary disks around HL Tau and AS 209. Our network predicts masses of $ 86.0  \pm 5.5 M_{\Earth} $, $ 43.8 \pm 3.3 M_{\Earth} $, and $ 92.2  \pm 5.1 M_{\Earth} $ in gaps at 10au, 30 au, and 80 au, respectively, in HL Tau. Similarly, for the AS 209 disk we find planet masses of $20.6 \pm 2.2 M_{\Earth} $ and  $ 36.3  \pm 2.5 M_{\Earth}$  at the 9 au and 100 au gaps, respectively. These estimates are in close agreement with results from other studies based on specialized simulations \citep{Dong2015OBSERVATIONALDISKS,Zhang2018TheInterpretation}

\end{itemize}

\section*{Acknowledgments}
SA and JBS acknowledge support from NASA under {\em Emerging Worlds} through grant 80NSSC20K0702.
DC acknowledges support from NASA under {\em Emerging Worlds} through grant 80NSSC21K0037. MKL is supported by the Ministry of Science and Technology of Taiwan (grants 107-2112-M-001-043-MY3, 110-2112-M-001-034-, 110-2124-M-002-012-) and an Academia Sinica Career Development Award (AS-CDA-110-M06). Numerical simulations were performed on the TIARA cluster at ASIAA, as well as the TWCC cluster at the National Center for High-performance Computing in Taiwan.

\newpage
\appendix 

\section{Steady-state drift solution for fixed particle size}\label{append1}

The current public version of the \textsc{fargo3d} code has a multi-fluid setup with pressureless dust species interacting with the gas by means of drag force parametrized by a constant Stokes number. We update this version of the code to incorporate a steady background solution for the radial drift of dust for fixed particle sizes instead of fixed Stokes number. Following the generalized steady-state solutions from the \cite{llambay19} for a vertically integrated disk with an isothermal equation of state, with pressure $P= c_{\rm s}^2 \Sigma$, we rewrite the exact background solutions for the gas and dust (i.e. in the limit of zero coupling) as 

\begin{equation}
\bm{v}_g = \beta(r)\bm{v}_K
\end{equation}
\begin{equation}
    \bm{v}_d = \bm{v}_K
\end{equation}
respectively, where $\bm{v}_K = r\Omega_K(r)\hat{\bm{\phi}}$ is the Keplerian velocity, $\hat{\bm{\phi}}$ is the unit vector in the azimuth, and 

\begin{equation}
\beta = \sqrt{1+2\eta(r)} , \eta = \frac{h^2}{2} \frac{d \log P }{ d \log r}  .  
\end{equation}
When dust and gas are coupled, we write their velocities as $\bm{v}_{d,g} \to \bm{v}_{d,g} + \delta\bm{v}_{d,g}$. Considering the steady-state axisymmetric Navier-Stokes equations for the perturbed velocities (see equations 63-66 in \cite{llambay19}), the gas and dust velocity perturbations in the radial direction is given as

\begin{equation} \label{deltavgr}
    \delta v_{gr} (r) = -2 \beta \mathcal{Q}_{N}\Psi (\beta -1 ) v_{\rm K},
\end{equation}

\begin{equation}
    \delta v_{dr} (r) = \frac{2S_t}{1+S_t^2}(\beta -1) v_{\rm K} + \frac{\delta v_{gr}+2S_t\delta v_{g\phi}}{1+ S_t^2}
\end{equation}
and the azimuthal counterpart is 

\begin{equation}
    \delta v_{g \phi} (r) = -\left[(\mathcal{S}_{N} + 2\xi) \mathcal{S}_{N} + \mathcal{Q}_{N}^2 \Psi (\beta - 1) v_{\rm K}\right]
\end{equation}

\begin{equation} \label{deltavdphi}
    \delta v_{d \phi} =  \frac{1}{1+S_t^2}(\beta -1) v_{\rm K} + \frac{2\delta v_{g\phi}-S_t\delta v_{gr}}{2(1+ S_t^2)}
\end{equation}

where 

\begin{equation} \label{SN}
\mathcal{S}_{N} \equiv \frac{\epsilon}{1+ S_t^2},  \mathcal{Q}_{N} \equiv \frac{\epsilon S_t}{1+S_t^2},    
\end{equation}

\begin{equation}\label{Psi}
    \Psi = \left[(\mathcal{S}_{N}  + \beta)(\mathcal{S}_{N} +2 \xi) + \mathcal{Q}_{N}^2 \right] \, \rm {and} \, \xi \equiv \beta \left(\frac{1}{2} + \frac{d log P }{ d log r} \right)
\end{equation}

Expanding equations \ref{deltavgr} to\ref{deltavdphi} using equations \ref{SN} and \ref{Psi} we get

\begin{equation} \label{vgr}
    \delta v_{gr} (r) = \frac{-2\beta (\beta-1)v_{\rm K} S_t \epsilon}{\epsilon^2 + \epsilon \beta(2\gamma + 3) + 2\beta^2 (\gamma +1) + 2 \beta^2 St^2(\gamma +1))}
\end{equation}

\begin{equation}
    \delta v_{g \phi } (r) = -\frac{(\beta -1) \epsilon v_{\rm K} (2 \beta \gamma + \epsilon + 2 \beta)}{ \epsilon^2 + 2 \beta^2 St^2 (\gamma + 1) + \beta \epsilon (2\gamma + 3) + 2 \beta^2(\gamma+1)}
\end{equation}

\begin{equation} \label{vdgr}
    \delta v_{d r} = \frac{4 \beta ^2 (\beta+ 1) S_{t} v_{\rm k}(\gamma +1 )}{\epsilon^2 + \epsilon \beta(2\gamma + 3) + 2 \beta^2 (\gamma +1) + 2 \beta^2 S_{t}^2 (\gamma +1)}
\end{equation}

\begin{equation}\label{vdphi}
    \delta v_{d \phi} = \frac{v_{\rm k}(\beta -1) (\epsilon \beta + 2 \beta^2(\gamma +1))}{\epsilon^2 + \epsilon \beta(2\gamma +3)+ 2\beta^2(\gamma+1) + 2 \beta^2 S_{\rm t}^2 (\gamma +1 )}
\end{equation}
The above solution Equations \ref{vgr} - \ref{vdphi} are solutions to steady-state drift solution only if they satisfy the following continuity equations

\begin{equation} \label{continuity}
    \partial_{r} (r\Sigma_{g}\delta v_{gr}) = 
     \partial_{r} (r\Sigma_{d}\delta v_{d}) = 0 .
\end{equation}
The particle size is parametrized by stokes number, $S_{\rm t}= t_{\rm s} \Omega_{\rm K}$, where the stopping time characterizing the strength of the dust-gas coupling is given as
\begin{equation}
    t_{s} = \rho_{\rm p}s/(\rho_{\rm g} c_{\rm s} ) = \frac{\pi s \rho_{\rm p}}{2 \Sigma_{\rm g}}.
\end{equation}
Here $\rho_{\rm p}$ is the density of the dust particles, $s$ is the radius of the dust particles and $\Sigma_{g}$ is the gas surface density. 
For fixed particle size $s = constant$ while the Stokes number $S_{\rm t} \propto \Sigma_{\rm g}^{-1}$ is a function of disk radius since $\Sigma_{\rm g} \propto r^{-\sigma}$ unlike the default setup where $S_{\rm t}$ is set to constant.

\begin{figure*}[ht!]
\centering
\includegraphics[scale=.80,]{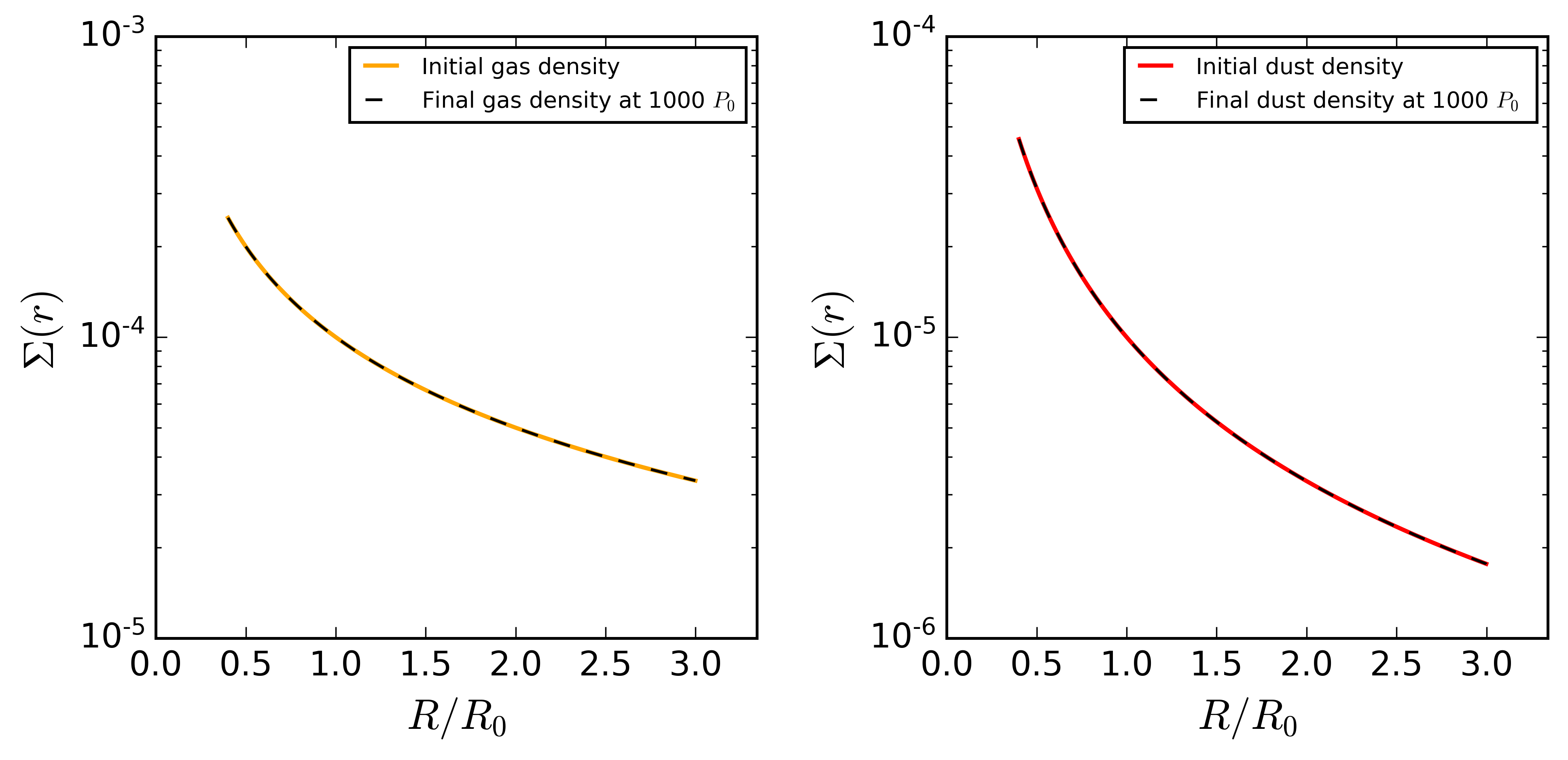}
\caption{The radial profile of the azimuthally averaged gas and dust surface density at the beginning of the simulation and after $1000 P_{\rm 0}$. The steady state profile is maintained with little evolution in the absence of an embedded planet. }\label{fig:steady-state}
\end{figure*}
To maintain a constant mass flux, i.e., $\partial_{r} (r\Sigma_{g}\delta v_{gr}) = \partial_{r} (r\Sigma_{d}\delta v_{d}) = 0$, we set 

\begin{equation}
    r\Sigma^0_{g}\delta v_{gr} = \rm {constant}
\end{equation}
Here we only consider the gas component but same result can be arrive for dust as well.
Furthermore, since $r\Sigma_{\rm g} \propto \left(\frac{r}{r}\right)^{1- \sigma} $, $\delta v_{\rm g r}$ must scale as 
\begin{equation}\label{vgr1}
    \delta v_{\rm g r} \propto \left(\frac{r}{r}\right)^{\sigma-1}
\end{equation}
such that the net mass flux is constant. Thus from equations \ref{vgr} and \ref{vgr1} we can establish

\begin{equation}\label{velradial_sol}
    \delta v_{\rm gr} (r) = \delta v_{\rm gr0} \left(\frac{r}{r}\right)^{\sigma-1} = \frac{a \epsilon}{\epsilon^2+b\epsilon+c}
\end{equation}
where $a = -2\beta (\beta-1)v_{\rm K} S_t \epsilon$,
$b= \beta (2\gamma + 1)$,
$c =  2\beta^2 (\gamma +1) + 2 \beta^2 St^2(\gamma +1) $ and $\delta v_{\rm gr0} = \frac{a_{\rm 0} \epsilon_{\rm 0}}{\epsilon_{\rm 0}^2+b_{\rm 0}\epsilon_{\rm 0}+c_{\rm 0}} $, and the subscript  ``0" refers to the values at $r = r_{0}$. 
Rearranging Equation \ref{velradial_sol} we solve for the dust-to-gas ratio $\epsilon$ and arrive at  

\begin{equation}\label{epsilon_new}
\epsilon = \frac{-(bd_{0}-a)\pm \sqrt{((b d_{\rm 0}-a)^2-4d_{\rm 0}^2c}}{2d_{\rm 0}},
\end{equation}
where $d_{\rm 0} = \frac{a_{\rm 0} \epsilon_{\rm 0}}{\epsilon_{\rm 0}^2+b_{\rm 0}\epsilon_{\rm 0}+c_{\rm 0}}  \left(\frac{r}{r}\right)^{\sigma-1} $ and the negative root gives a physical solution. Thus in order to satisfy the condition for constant mass flux for fixed particle size we adjusts the dust-to-gas ratio $\epsilon$ accordingly.

Furthermore, the viscous velocity of the gas $V^{\rm 2D}_{\rm vis}$ without dust grains is given as 
\begin{equation}\label{viscousv}
    V^{\rm 2D}_{\rm vis} = - \frac{3 \nu}{R} \frac{d \, \ln (\nu \Sigma_{\rm g} R^{1/2})}{d \, \ln R},
\end{equation}
\citep{lyn74}, where $\nu = \alpha c_{\rm s} h $ is the kinematic viscosity. For steady state background solution. i.e., $V^{\rm 2D}_{\rm vis}= 0$, we set $\nu \Sigma_{\rm g} R^{1/2} = \rm {constant}$ in Eq (\ref{viscousv}) which is equivalent to 

\begin{equation}\label{alpha}
  \alpha = \alpha_{\rm 0} (R/R_{\rm 0})^{\sigma-2F-1}.
\end{equation}

where the subscript ``0" refers to the values at $r = r_{0}$.

We initialize the disk-planet FARGO3D simulation with the updated dust-to-gas ratio Equation (\ref{epsilon_new}) and fix the particle size to a constant value using the fixed particle module. Figure (\ref{fig:steady-state}) shows the azimuthally average initial dust and gas radial surface density and the final surface density profile after the system has evolved $1000 P_{\rm 0} $. As demonstrated the steady state profile is maintained with little evolution is the absence of an embedded planet.

\bibliography{planet_bayesian}{}
\bibliographystyle{aasjournal}

\end{document}